\documentclass[reprint,amsmath,amssymb,aps,]{revtex4-2}

\usepackage{xcolor}
\usepackage{ulem}

\usepackage{graphicx}
\usepackage{dcolumn}
\usepackage{bm}

\newcommand{\be}{\begin{equation}} 
\newcommand{\ee}{\end{equation}}
\newcommand{\bea}{\begin{eqnarray}}
\newcommand{\eea}{\end{eqnarray}}
\newcommand{\bh}{{\bf{h}}}
\newcommand{\bc}{{\bf{c}}}
\newcommand{\bS}{{\bf{S}}}
\newcommand{\bcl}{{\bf{C}}}
\newcommand{\bW}{{\bf{W}}}

\begin{document}
	
%\preprint{APS/123-QED}
	
	\title{Phase behavior and percolation properties of the primitive model \\ of Laponite suspension. TPT of Wertheim with ISM reference system}
%	\thanks{A footnote to the article title}%
	
	\author{Y. V. Kalyuzhnyi}
	\affiliation{University of Ljubljana, Faculty of Chemistry and Chemical Technology, Ve\v cna pot 113, Ljubljana, Slovenia,
		\\Institute for Condensed Matter Physics, Svientsitskoho 1, 79011 Lviv, Ukraine}%Lines break automatically or can be forced with \\
%	\author{Second Author}%
%	\email{Second.Author@institution.edu}
%	\affiliation{Institute for Condensed Matter Physics, Svientsitskoho 1, 79011 Lviv, Ukraine	}%
	
%	\collaboration{MUSO Collaboration}%\noaffiliation
	
%	\author{Charlie Author}
%	\homepage{http://www.Second.institution.edu/~Charlie.Author}
%	\affiliation{
%		Second institution and/or address\\
%		This line break forced% with \\
%	}%
	
	\date{\today}% It is always \today, today,
	
\begin{abstract}
		
Computation of the properties of associative fluids with the particles highly 
anisotropic in shape, using multi-density perturbation theory of Wertheim,
has long been a challenge. We propose a simple and efficient scheme that allow us
to perform such computations. The scheme is based on a combination of
thermodynamic perturbation theory and the interaction site model approach for molecular 
fluids due to Chandler and Andersen. Our method is illustrated by its application to 
calculation of the phase diagram and percolation properties of a primitive model of 
Laponite suspension proposed recently.	
	
	\end{abstract}
	
	\maketitle

{\it Introduction.} -- 
Since Wertheim's pioneering work
 \cite{wertheim1984fluids1,wertheim1984fluids2,wertheim1986fluids1,wertheim1986fluids2,Wertheim1987} 
 significant progress has been made in the development
and application of the multi-density thermodynamic perturbation theory (TPT) for associative fluids. The theory and its 
extensions have been widely used to describe the properties of the fluid of small associative molecules and their 
mixtures, polymers, liquid crystals, surfactants, colloids and biological macromolecules (including proteins), 
etc. 
\cite{muller2001molecular,economou2002statistical,paricaud2002recent,tan2008recent,mccabe2010chapter,vega2016review,paduszynski2012thermodynamic,lira2022wertheim,pappu2023phase,shaahmadi2023group,vlachy2023protein}. 
A common feature of almost all these studies is that the particles of the
reference systems used there are spherical in shape. However, due to significant
progress made recently in the synthesis of colloidal building blocks of various 
shape and functionality \cite{glotzer2007anisotropy}, the possibility of a theoretical description of the effects of their self-assembling becomes highly relevant.

\begin{figure}
	\centering
	\includegraphics[width=0.48\textwidth]{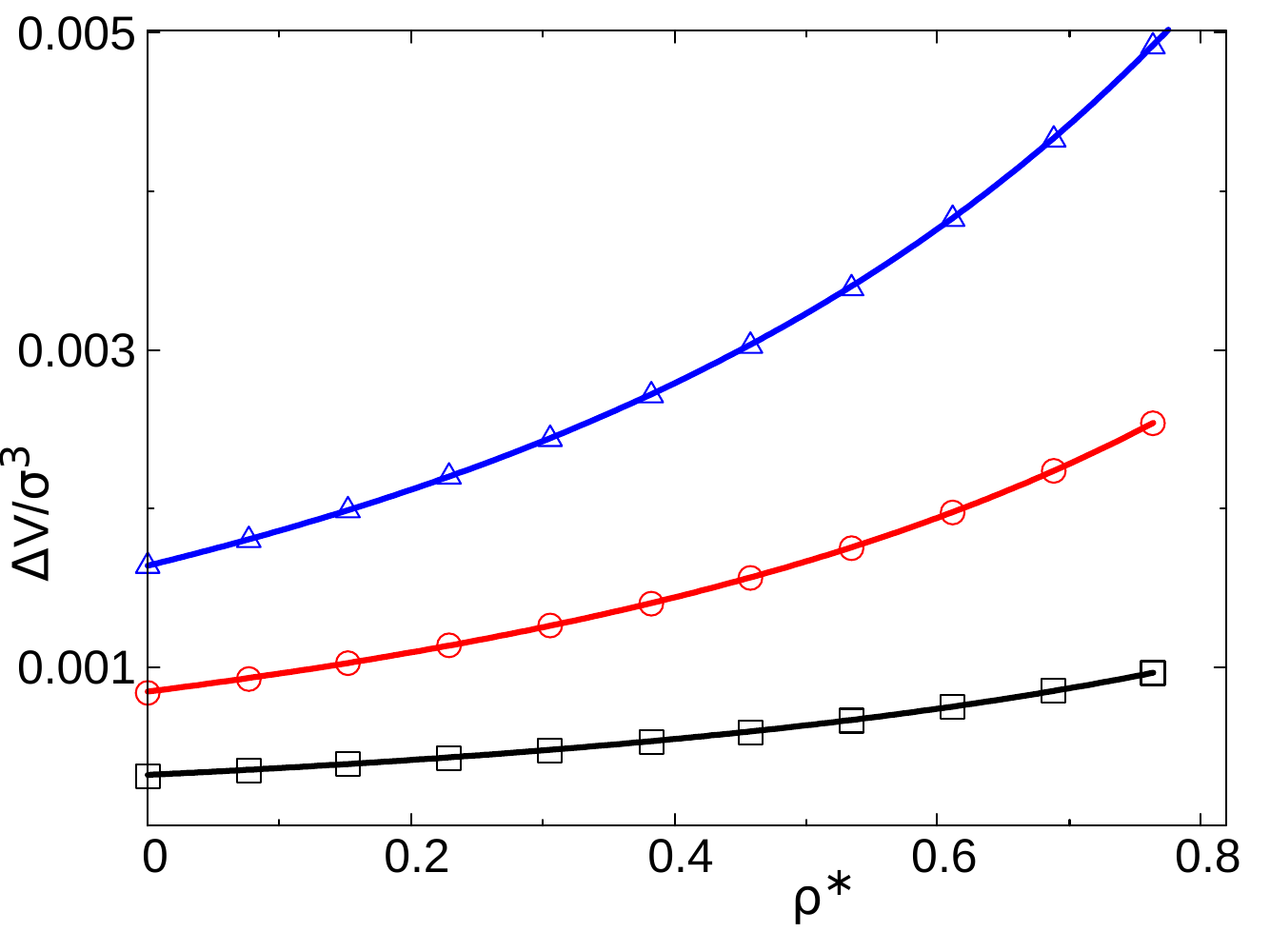}
	\caption{$\Delta V_{PY}$ (lines) and $\Delta V_{RISM-PY}$ (symbols) vs density $\rho^*$ at $d=0.5\sigma$ (black line and squares), $d=0.45\sigma$ (red line
	and circles) and $d=0.4\sigma$ (blue lines and triangles)}.
	\label{fig1}
\end{figure}

Initially, the TPT for associative fluids was formulated for a model represented 
by a fluid of hard spheres of size $\sigma$ with 
$n_s$ additional off-center square-well sites located at a distance $d\leq\sigma/2$ from the center of the hard sphere \cite{wertheim1986fluids1,wertheim1986fluids2,Wertheim1987,chapman1988phase}. 
Corresponding interparticle pair potential is
\be
U(12)=U_{hs}(r)+\sum_{KL} U_{KL}(12),
\label{U12}
\ee
where $U_{hs}(r)$ is the hard-sphere potential, $U_{KL}(12)$ is the site-site square-well potential acting between the
site $K$ of the particle 1 and site $L$ of the particle 2, i.e.
\be
U_{KL}(12)  = U_{KL}(z_{12})=
\left\{
\begin{array}{rl}
	\epsilon_{KL}, & z_{12}<\delta \\
	0, & z_{12}>\delta
\end{array}
\right.,
%\;\;\;\;z_{12}=|{\bf r}_2+{\bf d}_L(\Omega_1)-{\bf r}_1-{\bf d}_K(\Omega_2)|
\label{Uab1}
\ee
$z_{12}$ is the distance between sites $K$ and $L$, i.e.
$z_{12}=|{\bf r}_2+{\bf d}_L(\Omega_1)-{\bf r}_1-{\bf d}_K(\Omega_2)|$,
${\bf d}_K ({\bf d}_L)$ is the vector of the length $d$ connecting 
the center of the particle and its site $K(L)$,
1(2) denotes position ${\bf r}_1({\bf r}_2)$ and orientation $\Omega_1(\Omega_2)$ of the particle 1(2). 
Here $K$ and $L$ take $n_s$ values $A,B,C,\ldots$.
The parameters of the square-well site-site potential $d$ and $\delta$
were chosen to satisfy the 'one bond per site' condition
$\delta<\sqrt{\sigma^2+d^2-\sigma d\sqrt{3}}-d$
, i.e. each site of one 
particle can be involved in a bond with only one site of another particle. 
The first-order version of the TPT (TPT1) is formulated in terms of the 
Helmholts free energy $A$ of the model, which is represented as the sum of two terms, i.e.
\be
A=A_{ref}+\Delta A_{as},
\label{A}
\ee
where $A_{ref}$ is Helmholtz free energy of the reference system and $\Delta A_{as}$ is contribution to Helmholtz free 
energy due to association,
\be
\beta{\Delta A_{as}\over N}=\sum_{K}\left(\ln{X_K}-{1\over 2}X_K\right)+{1\over 2}n_s.
\label{DAas}
\ee
Here $\beta=1/(k_BT)$, $k_B$ is Boltzmann’s constant, $T$ is temperature, $N$ is the number of the particles,
$X_K$ is fraction of the particles with attractive site of the type $K$ not bonded. 
This fraction follows from the solution of the set of equations 
\be
\rho X_K\sum_L X_LI_{KL}+X_K-1=0,
\label{mass}
\ee
where
\be
I_{KL}=\int\langle g_{ref}(12)f_{KL}(12)\rangle_{\Omega_1\Omega_2} \; d{\bf r}_{12}.
\label{I}
\ee
Here $\rho$ is the number density of the system,
$g_{ref}(12)$ is the pair distribution function of the reference system, $f_{KL}(12)$ is
the Mayer function for the site-site square-well potential, i.e $f_{KL}(12)=\exp{[-\beta U_{KL}(12)]}-1$, and
$\langle\ldots\rangle_{\Omega_1\Omega_2}$ denotes angular averaging with respect to orientations of particles 1 and 2.
This integral can be calculated assuming any location of the origin of the coordinate 
system that is associated with the particle. Assuming that the location of the origin 
coincides with the location of the corresponding attractive site of the particle (site 
$K$ of the particle 1 and site $L$ of the particle 2) we have 
\be
I_{KL}=4\pi\int\langle g_{ref}(12)\rangle_{\Omega_1\Omega_2}r_{12}^2f_{KL}(r_{12})\;dr_{12},
\label{Iaux1}
\ee
where $r_{12}$ is the distance between sites $K$ and $L$ of particles 1 and 2, respectively, and 
$\langle g_{ref}(12)\rangle_{\Omega_1\Omega_2}$ is the site-site pair distribution function between two auxiliary
sites $K$ and $L$ of the reference system 
\cite{chandler1973derivation,cummings1981auxiliary}. For the model at hand displacement of the sites from
the hard-sphere center $d$ is the same for each site, therefore corresponding site-site distribution function,
which we will denote as $g^{(ref)}_{ss}(r)$, 
do not depend on the type of sites, i.e. $\langle g_{ref}(12)\rangle_{\Omega_1\Omega_2}=g^{(ref)}_{KL}(r)=g^{(ref)}_{ss}(r)$. 
This correlation function can be calculated either using reference
interaction site model (RISM) approach due to Chandler \cite{chandler1972optimized,monson1990recent} or performing direct averaging using the 
appropriate expression for hard-sphere radial distribution function. For the model at hand RISM approach is 
represented by the site-site Ornstein-Zernike (SSOZ) equation
\be
{\hat \bh}(k)={\hat \bS}(k){\hat \bc}(k){\hat \bS}(k)+\rho{\hat \bS}(k){\hat \bc}(k){\hat \bh}(k),
\label{SSOZ}
\ee
and Percus-Yevick-like closure relation
\be 
\left\{
\begin{array}{rl}
c_{\alpha\beta}(r)  =\;\;\;	0, & r>\sigma-d\Delta_{\alpha\beta}
\\
h_{\alpha\beta}(r)	= -1, & r\leq \sigma-d\Delta_{\alpha\beta}
\end{array}
\right.,
\label{closure1}
\ee
where $\Delta_{\alpha\beta}=2\delta_{\alpha s}\delta_{\beta s}+
\delta_{\alpha 0}\delta_{\beta s}+\delta_{\alpha s}\delta_{\beta 0}$, $\alpha$ and $\beta$ take the values
0 and $s$ where 0 denotes the center of the particle and $s$ its off-center site, 
${\hat \bS}(k)$ is a matrix with elements 
$S_{\alpha\beta}(k)=\delta_{\alpha\beta}+(1-\delta_{\alpha\beta})\sin{(kd)}/(kd)$,
${\hat \bh}(k)$ and ${\hat \bc}(k)$ are matrices with elements 
${\hat h}_{\alpha\beta}(k)$ and ${\hat c}_{\alpha\beta}(k)$, which are Fourier transforms of the total and
direct site-site correlation functions $h_{\alpha\beta}(r)$ and $c_{\alpha\beta}(r)$, respectively, and
$\delta_{\alpha\beta}$ is Kroneker delta. Alternatively we have \cite{chandler1972optimized,holovko1990electrostatic}
\be
g^{(ref)}_{ss}(r)={1\over 4d^2r}\int_{|r-d|}^{r+d}dt\int_{|t-d|}^{t+d}\upsilon g_{hs}(\upsilon)\;d\upsilon,
\label{direct}
\ee
where $g_{hs}(r)$ is the radial distribution function of hard spheres. Using Percus-Yevick expression 
for $g_{hs}(r)$ \cite{smith1970analytical}
, we have
\be
g^{(ref)}_{ss}(r)={\sigma^3\over 4d^2r}\sum_{i=0}^2{a_i\over t_i}\left[{1\over t_i}\left(e^{r_dt_i}-1\right)-r_d\right],
\label{zone}
\ee
where $r\le 2(\sigma-d)$, $r_d=(r+2d-\sigma)/\sigma$ and $a_i=t_iL(t_i)/ S_1(t_i),$ 
\be
t_i={-2\eta+(y_+j^i+y_-j^{-i})\sqrt[3]{2\eta \xi}\over 1-\eta},
\label{coef1}
\ee
\be
y_{\pm}=\sqrt[3]{1\pm\sqrt{1+2\left(\eta^2/ \xi\right)^2}},
\label{coef}
\ee
 $\xi=3+3\eta-\eta^2$,
$S_1(t)=3(1-\eta)^2t^2+12\eta(1-\eta)t+18\eta^2$,
$L(t)=(1+\eta/2)t+1+2\eta$,
$j=\exp{\left(2\pi\sqrt{-1}/3\right)}$. Corresponding expression for the integral $I_{KL}$ is
\be
I_{KL}=\left(e^{-\beta\epsilon_{KL}}-1\right)\Delta V_{PY},
\label{IKL}
\ee
where
\be
\Delta V_{PY}={\pi\over 6}\sum_{i=0}^2{a_i\over t_i^4d^2}
\left[6\left(\delta t_i-\sigma\right)e^{{2d+\delta\over\sigma}-1}-
{1\over\sigma^2}\sum_{l=0}^3P_lt_i^l\right],
\label{dV}
\ee 
$P_3=-8d^3+12d^2\sigma+6(\delta^2-\sigma^2)d+(2\delta-3\sigma)\delta^2+\sigma^3$, $P_2=3(-4d^2\sigma+4d\sigma^2+\delta^2\sigma-\sigma^3)$,
$P_1=6\sigma^2(\sigma-2d)$, $P_0=-6\sigma^3$. 
As expected, expression (\ref{IKL}) coincide with corresponding expression 
for $I_{KL}$, derived following the scheme suggested
by Wertheim \cite{wertheim1986fluids}, i.e. when the origin of the coordinate system of the particle is 
placed in the hard-sphere center. 
In figure \ref{fig1} we compare our results for $\Delta V_{PY}$ (\ref{dV}) and $\Delta V_{RISM-PY}$
as a function of density at three different values of $d$, i.e. $d=0.5,\;0.45,\;0.4$. Here
$\Delta V_{RISM-PY}$ is obtained using $g_{ss}(r)$ calculated by numerical solution of the RISM
equation (\ref{SSOZ}) with PY-like closure relations (\ref{closure1}). Excellent agreement 
is observed, i.e. on the scale of the figure the results for $\Delta V_{PY}$ and $\Delta V_{RISM-PY}$ 
coincide. Thus, for models with hard-sphere reference system, the use of either of the two 
methods will give practically the same results. In this case, the scheme suggested by Wertheim has the 
advantage of being simpler and easier to use. However, for models with non-spherical particles, the calculation 
of the key integral $I_{KL}$ within this scheme is a formidable task. 
On the other hand, this integral can be calculated relatively easily using a method that 
uses site-site distribution functions, especially in the case when the structure of the reference system 
can be described within the framework of the interaction site formalism of Chandler et al. 
\cite{chandler1972optimized}. In addition to the fluids of small molecules 
\cite{,monson1990recent,hansen2013theory,gray2011theory} models of this 
type are widely used to describe the properties of macromolecular and colloidal fluids 
\cite{zhang2003tethered,zhang2004self,costa2005structure,glotzer2007anisotropy,delhorme2012monte}.

In this Letter, we illustrate the application of the proposed  scheme by presenting our calculations for
the phase behavior of a primitive model of Laponite suspension. The model and corresponding 
computer simulation studies of its liquid-gas phase behavior were recently presented by 
Ruzicka et al. \cite{ruzicka2011observation}. 

\begin{figure}
\centering
\includegraphics[width=0.38\textwidth]{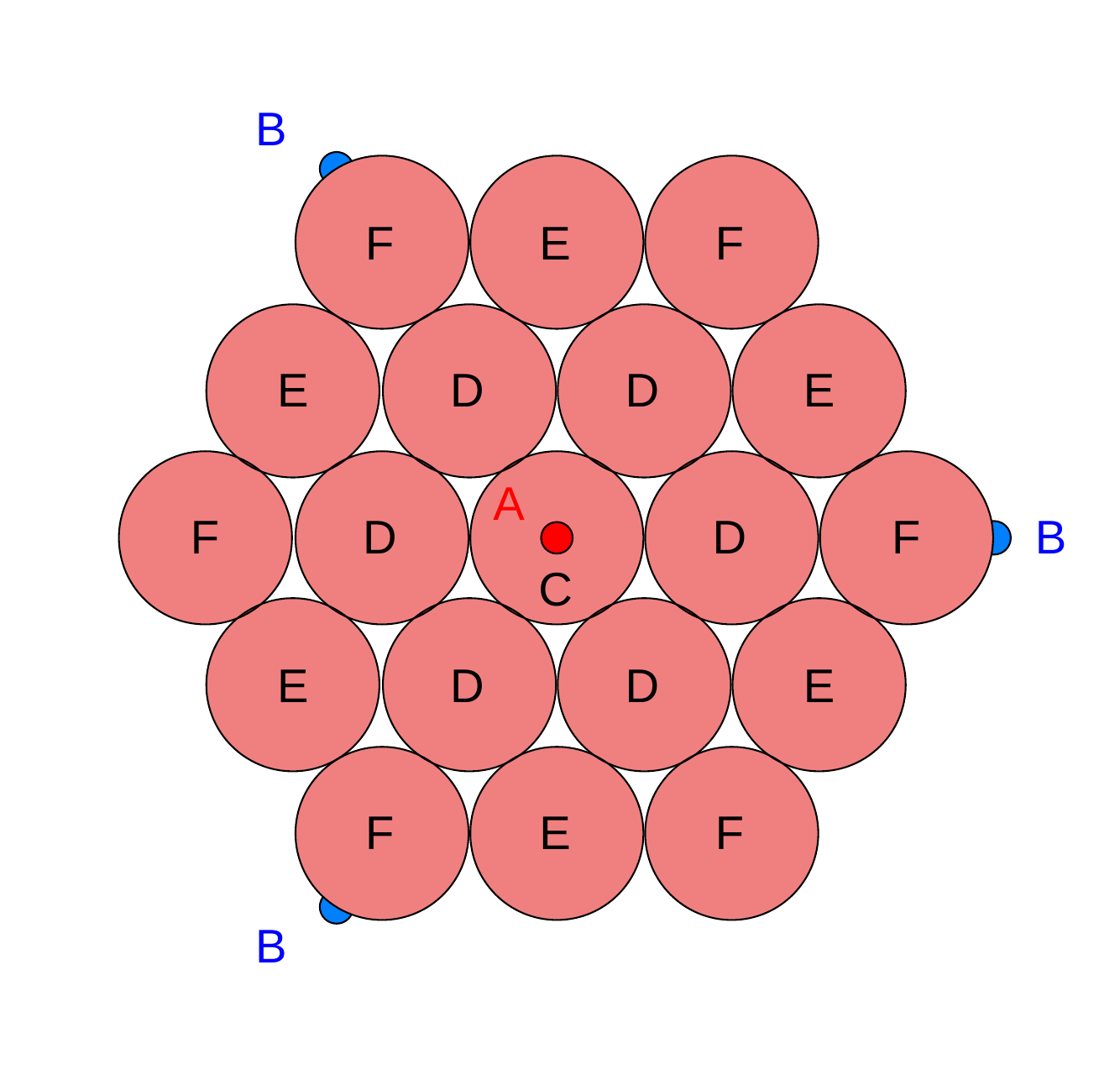}
\caption{Schematic representation of the primitive model of Laponite nanoparticles.
Here square-well sites are denoted as $A$ (red) and $B$ (blue) and hard-sphere sites
are denoted as $C,D,E,F$ (black). }
\label{fig2}
\end{figure}

{\it Primitive model of Laponite.} -- The primitive model of a Laponite nanoparticle is represented
by a collection of 19 hard spheres of size $\sigma$ arranged to form a hexagonal shaped platelike particle, with their
nearest neighbors in contact. In addition, three symmetrically located at the corners of the hexagon 
hard spheres are decorated with one square-well site of type $B$ each, and the central hard sphere of the
hexagon is decorated with two square-well sites of type $A$, placed on its two opposite
faces (see figure \ref{fig2}). The number of sites of type $A$ is $n_A=2$ and of  
type $B$ is $n_B=3$. 
%Thus the valency of the model (the number of the bonding sites per particle) 
%is $n_s=n_A+n_B=5$. 
Site-site square-well interaction is 
acting only between sites of different type, i.e.
$\epsilon_{KL}=(1-\delta_{KL})\epsilon$. These sites are placed on the surface of the respective 
hard sphere, so $d=\sigma/2$. The width of the square-well $\delta=0.1197\sigma$

{\it Theory.} -- Taking into account the symmetry of the model and using the expression for the Helmholtz free energy (\ref{DAas}), we have
\be
{\beta \Delta A_{as}\over V}=\rho\left[\ln{(X_A^2X_B^3)}-{1\over 2}\left(2X_A+3X_B\right)
+{5\over 2}\right],
\label{DAas1}
\ee
where fractions $X_A$ and $X_B$ follow from the ``mass action law'' equation (\ref{mass}), i.e.
\be
X_A={-1-\rho I_{AB}+\sqrt{(1+\rho I_{AB})^2+8\rho I_{AB}}\over 2\rho I_{AB}}
\label{XA2}
\ee
and
\be
X_B={1\over 3}(1+2X_A).
\label{XA1}
\ee
Expression for the integral $I_{AB}$ is
\be
I_{AB}=4\pi(e^{-\beta\epsilon}-1)\int_0^\delta r_{12}^2g^{(ref)}_{AB}(r_{12})\;dr_{12},
\label{IAB}
\ee
where $g^{(ref)}_{AB}(r_{12})$ is the site-site pair distribution function of the reference system 
between the sites $A$ and $B$. All other thermodynamic properties follow from standard
thermodynamical relations. For pressure $P$ and chemical potential $\mu$ we have
\be
\beta P=\rho+\beta P_{ref}^{(ex)}+\beta \Delta P_{as},
\label{t-d1}
\ee
\be
\beta\mu=\ln{(\rho\Lambda^3)}+\beta \mu^{(ex)}_{ref}+\beta\Delta\mu_{as},
\label{t-d}
\ee
where $P_{ref}^{(ex)}$ and $\mu_{ref}^{(ref)}$ are excess pressure and chemical potential potential, respectively.
For the contributions to chemical
potential $\Delta\mu_{as}$ and pressure $\Delta P_{as}$ due to association we have
\be
\Delta\mu_{as}=\left({\partial(\Delta A_{as}/V)\over\partial\rho}\right)_{T,V},
\label{thermo1}
\ee
\be
\Delta P_{as}=\rho\Delta\mu_{as}-\Delta A_{as}/V.
\label{thermo}
\ee 

{\it Properties of the reference system.} -- The reference system is represented by a fluid of 
Laponite particles with zero site-site square-well depth, i.e. $\epsilon=0$. Both the
thermodynamic and structural properties of such reference system are calculated using 
the appropriate 
versions of the SSOZ equation, supplemented by a PY-like closure. 
The thermodynamics of the reference system does not depend on the presence or absence of auxiliary
sites, therefore we consider the model with hard-sphere sites only. There are 19 hard-sphere sites,
thus the dimension of the matrices representing site-site correlation functions in 
the SSOZ
equation is $19\times 19$. However, taking into account the symmetry of the model, the dimensionality
of the SSOZ equation can be reduced. We follow here the scheme proposed by Raineri and Stell
\cite{raineri2001dielectrically} and recently used by Costa et al. \cite{costa2005structure} 
to study a model similar to the current one.
We identify 4 groups of equivalent hard-sphere sites of the model, which we denote as $C,D,E$ 
and $F$ (see figure \ref{fig2}). Each group of the type $K$ includes $n_K$ sites denoted as
$K_1,K_2,\ldots,K_{n_K}$. For this model $n_C=1$ and $n_D=n_E=n_F=6$. Now reduced version
of the SSOZ equation can be written as follows
\be
{\hat \bh}(k)={\hat \bW}(k){\hat \bcl}(k){\hat \bW}(k)+\rho{\hat \bW}(k){\hat \bcl}(k){\hat \bh}(k),
\label{redSSOZ}
\ee
where ${\hat \bW}(k)$ and ${\hat \bcl}(k)$ are matrices 
with elements
\be
{\hat W}_{KL}(k)={1\over n_L}\sum_{j=1}^{n_L}{\hat S}_{ij}(k)=
{1\over n_K}\sum_{i=1}^{n_K}{\hat S}_{ij}(k)={\hat W}_{LK}(k)
\label{W}
\ee
and ${\hat C}_{KL}(k)=n_K{\hat c}_{ij}(k)n_L$. The corresponding PY-like closure is
\be 
\left\{
\begin{array}{rl}
	C_{KL}(r)  =\;\;\;	0, & r>\sigma
	\\
	h_{KL}(r)	= -1, & r\leq \sigma
\end{array}
\right..
\label{closure2}
\ee
The solution of this set of equations is used to calculate the thermodynamic properties of the model
using compressibility rout. The corresponding expressions for the excess values of the pressure 
$P^{(ex)}_{ref}$ and chemical potential $\mu^{(ex)}_{ref}$ are
\be
\beta P^{(ex)}_{ref}=-4\pi\int_0^\rho\rho'\;d\rho'\sum_{KL}\int r^2C_{KL}(r)\;dr
\label{Pex}
\ee
and
\be
\beta\mu^{(ex)}_{ref}=-4\pi\int_0^\rho\;d\rho'\sum_{KL}\int r^2C_{KL}(r)\;dr.
\label{muex}
\ee

The calculation of the structure properties requires the solution of the SSOZ equation formulated for
a model that in addition to hard-sphere sites includes also auxiliary sites. We consider 
a model with eight auxiliary sites. Six of the sites represent square-well sites with $\epsilon=0$, while the remaining three
are introduced to increase the degree of symmetry of the model. The last three 
sites are placed on the surface of three rim hard-sphere sites that are not decorated 
with square-well sites (see figure \ref{fig2}). Thus, there are six groups of equivalent sites, i.e.
$A,\;B,\;C,\;D,\;E,\;F$, where the first two represent auxiliary sites and the last 
four
represent hard-sphere sites. Here $n_A=2$ and $n_B=6$. Now the dimension of the matrices that 
enter the SSOZ equation (\ref{redSSOZ}) is $6\times 6$. The solution of this version of SSOZ equation gives
$g_{AB}^{(ref)}(r)$, which is used to calculate the integral $I_{AB}$ (\ref{IAB}).

{\it Results and discussion.} -- Using the theory developed above, we calculate the liquid-gas
phase diagram and the percolation threshold line of the primitive model of Laponite suspension.
The densities of the coexisting phases follow from the solution of the set of equations 
representing the phase equilibrium conditions
\be
\left\{
\begin{array}{ll}
P(T,\rho_{g})=P(T,\rho_l)&
\\
\mu(T,\rho_{g})=\mu(T,\rho_l)&
\end{array}
\right.,
\ee
where $\rho_g$ and $\rho_l$ are the densities of low-density and high-density phases, respectively.
The percolation thrshold line was calculated following the scheme suggested by Tavares et al. 
\cite{tavares2010percolation}. For a detailed description of the scheme, we refer readers to the 
original publication; here we present only the final set of equations to be solved.
The threshold line points on the $\rho$ vs $T$ coordinate plane satisfy
the following equation
\be
%\sqrt{\left(T_A+T_B\right)^2-4T_AT_B\left({1-X_A\over T_A}+{1-X_B\over T_B}
%-{1\over n_An_B}\right)}+T_A+T_B-2=0,
\sqrt{T_\Sigma^2-4T_\Pi\left({1-X_A\over T_A}+{1-X_B\over T_B}
	-{1\over n_\Pi}\right)}+T_\Sigma-2=0,
\label{perc}
\ee
where $T_L=n_L\left(1-X_L\right)\prod_{K=A}^Bq_K^{n_K-1}$, 
$T_\Pi=T_AT_B$, $n_\Pi=n_An_B$, $T_\Sigma=T_A+T_B$
and $q_L$ is 
obtained from the solution of the set of equations
\be
X_L-\left[1-\left(1-X_L\right)\prod_{K=A}^Bq_K^{n_K-1}\right]q_L=0.
\label{perc1}
\ee

\begin{figure}
	\centering
	\includegraphics[width=0.48\textwidth]{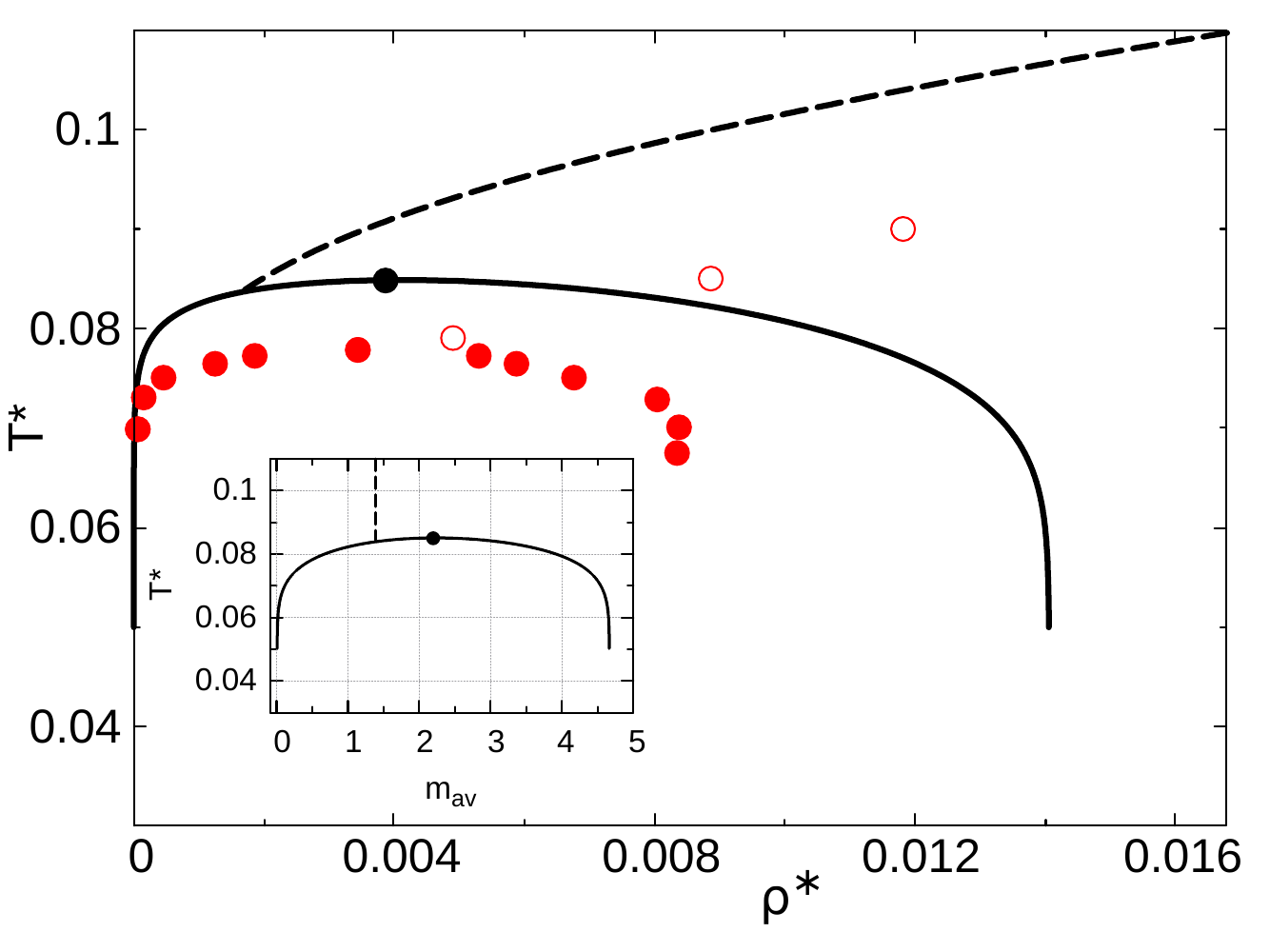}
\caption{Phase diagram and percolation threshold line of the primitive model of Laponit 
suspension in the $T^*$ vs $\rho^*$ coordinate frame. Here, the black lines and filled black circles 
represent theoretical results, the red circles represent computer simulation results 
\cite{ruzicka2011observation} and the filled black circles mark the critical point. 
The percolation threshold line is marked by black lines.
The inset in the figure shows the phase diagram and percolation threshold line in $T^*$ vs $m_{av}$ coordinate frame.}
	\label{fig3}
\end{figure}

In figure \ref{fig3} we present the theoretical and Monte Carlo computer simulation results 
\cite{ruzicka2011observation} for the phase diagram and the percolation threshold line using
$T^*$ vs $\rho^*$ coordinate frame. Here $T^*=k_BT/\epsilon$ and $\rho^*=\rho\sigma^3$,
where $k_B$ is the Boltzmann constant. 
In general the accuracy of the current version of Wetheim's first order TPT is similar 
to that observed in the case of hard spheres with several off-center square-well sites 
\cite{bianchi2008theoretical}. In the latter case, the theoretical predictions are only 
semiquantitatively accurate. While our theory gives relatively accurate predictions for the 
critical density $\rho^*_c$, results for the critical temperature $T_c^*$ and, therefore, for
percolation threshold line are less accurate being about 9$\%$ too large for $T_c^*$. In addition, the liquid branch of the phase diagram at low temperatures is located at
densities that are about 1.7 times too high. The width of the phase diagram and the
position of its liquid branch are determined by the effective valency of the model $\upsilon_{eff}$ , i.e. the average number 
of bonds per particle formed when the limit of infinitely low temperature is reached 
\cite{ruzicka2011observation}. As this number increases, the width of the phase diagram increases 
and its liquid branch moves towardsm the higher densities \cite{bianchi2008theoretical}. In the 
framework of the TPT1 the average number of the bonds per particle $m_{av}$ can be calculated 
 using the following expression
\be
m_{av}=\sum_mm\sum_{m_A+m_B=m}\prod_{L=A}^B{n_L!X_L^{\Delta n_L}\left(1-X_L\right)^{m_L}
	\over m_L!\left(\Delta n_L\right)!},
\label{mav}
\ee
where $m_A$ and $m_B$ take the values $0,1,2$ and $0,1,2,3$, respectively, and
$\Delta n_L=n_L-m_L$. According to (\ref{XA2}) and (\ref{XA1})
$\lim_{T^*\rightarrow 0}{X_A}=0$ and $\lim_{T^*\rightarrow 0}{X_B}=1/3$.
Thus at infinity low temperature all sites of the type $A$ are bonded. Using this result and 
expression for $m_{av}$ (\ref{mav}) we have
\be
\upsilon^{(TPT)}_{eff}=\lim_{T^*\rightarrow 0}m_{av}={14\over 3}.
\label{eff}
\ee 
This value of the effective valency defines the position of the liquid branch of the theoretical 
phase diagram at low temperatures. In the inset of figure \ref{fig3} we present the value of 
$m_{av}$ along
coexisting lines. Here it is seen that for temperatures below $\approx 0.065$ $m_{av}$
does not change much and approaches its limiting value with decreasing temperature. However,
according to MC computer simulation study \cite{ruzicka2011observation} exact value of
the effective valency is lower, i.e. $\upsilon_{eff}^{(MC)}=4$ and therefore the liquid branch of
the Monte Carlo phase diagram is located at a densities smaller than those of the theoretical 
phase diagram. Thus even at infinitely low temperature $X_A\ne 0$, i.e. there is certain 
fraction of the particles with site $A$ nonbonded. This behavior is due to the effects of 
blocking, i.e. when bonding of one site completely blocks bonding of the other. 
%Due to highly anisotropic shape of the current model these effects appears to be substantial. 
Unfortunately, TPT1 does not take into account blocking effects \cite{Wertheim1987}, 
which due to highly anisotropic shape of the model in question appear to be 
substantial.
Thus, this discrepancy between the exact and theoretical results for the location of 
the liquid branch of the phase diagram is caused by the failure to account for 
blocking effects within the TPT1 framework.
%
%Thus disagreement between exact and theoretical results for the location of the 
%liquid branch of the phase diagram is caused by the inappropriate account of the 
%blocking effects in the framework of TPT1. 
Nevertheless general conclusion reached on 
the basis of the current study is similar to that obtained using MC simulation 
approach, i.e. the model proposed enables us to qualitatively correct describe formation of the empty liquid state, observed experimentally in Laponite
suspensions \cite{ruzicka2011observation,ruzicka2011fresh}.

In summary, we propose a novel theoretical scheme that allows us to efficiently apply Wertheim's multidensity TPT to study the properties of associating fluids with the particles of highly non-spherical shape. The scheme is based on the 
combination of the TPT and ISM formalism for molecular fluids, with the latter being used to calculate the 
properties of the reference system. For the models with the reference system described by the 
fluid of fused hard spheres the ISM approach is represented by RISM integral equation theory 
for site-site fluids. We expect that our results will boost research focused on the
theoretical description of self-assembling of particles of anisotropic shape.

We acknowledge financial support through the MSCA4Ukraine project 
(ID: 101101923), which is funded by the European Union.

\bibliography{EmptyPrl}

@PREAMBLE{
 "\providecommand{\noopsort}[1]{}" 
 # "\providecommand{\singleletter}[1]{#1}%" 
}

@article{wertheim1984fluids1,
  title={Fluids with highly directional attractive forces. I. Statistical thermodynamics},
  author={Wertheim, Michael S},
  journal={Journal of statistical physics},
  volume={35},
  number={1},
  pages={19--34},
  year={1984},
  publisher={Springer}
}

@article{wertheim1984fluids2,
  title={Fluids with highly directional attractive forces. II. Thermodynamic perturbation theory and integral equations},
  author={Wertheim, Michael S},
  journal={Journal of statistical physics},
  volume={35},
  number={1},
  pages={35--47},
  year={1984},
  publisher={Springer}
}

@article{wertheim1986fluids1,
  title={Fluids with highly directional attractive forces. III. Multiple attraction sites},
  author={Wertheim, Michael S},
  journal={Journal of statistical physics},
  volume={42},
  number={3},
  pages={459--476},
  year={1986},
  publisher={Springer}
}

@article{wertheim1986fluids2,
  title={Fluids with highly directional attractive forces. IV. Equilibrium polymerization},
  author={Wertheim, Michael S},
  journal={Journal of statistical physics},
  volume={42},
  number={3},
  pages={477--492},
  year={1986},
  publisher={Springer}
}

@article{Wertheim1987,
  title={Thermodynamic perturbation theory of polymerization},
  author={Wertheim, M S},
  journal={J. Chem. Phys.},
  volume={87},
  number={12},
  pages={7323--7331},
  year={1987},
  publisher={American Institute of Physics}
}

@article{muller2001molecular,
  title={Molecular-based equations of state for associating fluids: A review of SAFT and related approaches},
  author={M{\"u}ller, Erich A and Gubbins, Keith E},
  journal={Industrial \& engineering chemistry research},
  volume={40},
  number={10},
  pages={2193--2211},
  year={2001},
  publisher={ACS Publications}
}

@article{economou2002statistical,
  title={Statistical associating fluid theory: A successful model for the calculation of thermodynamic and phase equilibrium properties of complex fluid mixtures},
  author={Economou, Ioannis G},
  journal={Industrial \& engineering chemistry research},
  volume={41},
  number={5},
  pages={953--962},
  year={2002},
  publisher={ACS Publications}
}

@article{paricaud2002recent,
  title={Recent advances in the use of the SAFT approach in describing electrolytes, interfaces, liquid crystals and polymers},
  author={Paricaud, Patrice and Galindo, Amparo and Jackson, George},
  journal={Fluid phase equilibria},
  volume={194},
  pages={87--96},
  year={2002},
  publisher={Elsevier}
}

@article{tan2008recent,
  title={Recent advances and applications of statistical associating fluid theory},
  author={Tan, Sugata P and Adidharma, Hertanto and Radosz, Maciej},
  journal={Industrial \& Engineering Chemistry Research},
  volume={47},
  number={21},
  pages={8063--8082},
  year={2008},
  publisher={ACS Publications}
}

@incollection{mccabe2010chapter,
  author    = {McCabe, C and Galindo, A},
  title     = {SAFT Associating fluids and fluid mixtures},
  booktitle = {Applied Thermodynamics of Fluids},
  chapter   = {8},
  pages     = {215–279},
  publisher = {RSC Publishing},
  year      = {2010},
  editor    = {A.R.H Goodwin and J.V Sengers},
}

@article{vega2016review,
  title={Review and new insights into the application of molecular-based equations of state to water and aqueous solutions},
  author={Vega, LF and Llovell, F},
  journal={Fluid Phase Equilibria},
  volume={416},
  pages={150--173},
  year={2016},
  publisher={Elsevier}
}

@article{paduszynski2012thermodynamic,
  title={Thermodynamic modeling of ionic liquid systems: development and detailed overview of novel methodology based on the PC-SAFT},
  author={Paduszynski, Kamil and Domanska, Urszula},
  journal={The Journal of Physical Chemistry B},
  volume={116},
  number={16},
  pages={5002--5018},
  year={2012},
  publisher={ACS Publications}
}

@article{lira2022wertheim,
  title={Wertheim’s association theory for phase equilibrium modeling in chemical engineering practice},
  author={Lira, Carl T and Elliott, J Richard and Gupta, Sumnesh and Chapman, Walter G},
  journal={Industrial \& Engineering Chemistry Research},
  volume={61},
  number={42},
  pages={15678--15713},
  year={2022},
  publisher={ACS Publications}
}

@article{pappu2023phase,
  title={Phase transitions of associative biomacromolecules},
  author={Pappu, Rohit V and Cohen, Samuel R and Dar, Furqan and Farag, Mina and Kar, Mrityunjoy},
  journal={Chemical reviews},
  volume={123},
  number={14},
  pages={8945--8987},
  year={2023},
  publisher={ACS Publications}
}

@article{shaahmadi2023group,
  title={Group-contribution SAFT equations of state: A review},
  author={Shaahmadi, Fariborz and Smith, Sonja AM and Schwarz, Cara E and Burger, Andries J and Cripwell, Jamie T},
  journal={Fluid Phase Equilibria},
  volume={565},
  pages={113674},
  year={2023},
  publisher={Elsevier}
}

@article{vlachy2023protein,
  title={Protein association in solution: Statistical mechanical modeling},
  author={Vlachy, Vojko and Kalyuzhnyi, Yurij V and Hribar-Lee, Barbara and Dill, Ken A},
  journal={Biomolecules},
  volume={13},
  number={12},
  pages={1703},
  year={2023},
  publisher={MDPI}
}

@article{glotzer2007anisotropy,
  title={Anisotropy of building blocks and their assembly into complex structures},
  author={Glotzer, Sharon C and Solomon, Michael J},
  journal={Nature materials},
  volume={6},
  number={8},
  pages={557--562},
  year={2007},
  publisher={Nature Publishing Group UK London}
}

@article{chapman1988phase,
  title={Phase equilibria of associating fluids: chain molecules with multiple bonding sites},
  author={Chapman, Walter G and Jackson, George and Gubbins, Keith E},
  journal={Molecular Physics},
  volume={65},
  number={5},
  pages={1057--1079},
  year={1988},
  publisher={Taylor \& Francis}
}

@article{chandler1973derivation,
  title={Derivation of an integral equation for pair correlation functions in molecular fluids},
  author={Chandler, David},
  journal={Journal of Chemical Physics},
  volume={59},
  number={5},
  pages={2742--2746},
  year={1973}
}

@article{cummings1981auxiliary,
  title={Auxiliary sites in the RISM approximation for molecular fluids},
  author={Cummings, PT and Gray, CG and Sullivan, DE},
  journal={Journal of Physics A: Mathematical and General},
  volume={14},
  number={6},
  pages={1483},
  year={1981},
  publisher={IOP Publishing}
}

@article{chandler1972optimized,
  title={Optimized cluster expansions for classical fluids. II. Theory of molecular liquids},
  author={Chandler, David and Andersen, Hans C},
  journal={The Journal of Chemical Physics},
  volume={57},
  number={5},
  pages={1930--1937},
  year={1972},
  publisher={American Institute of Physics}
}

@article{monson1990recent,
  title={Recent progress in the statistical mechanical mechanics of interaction site fluids},
  author={Monson, PA and Morriss, GP},
  journal={Advances in chemical physics},
  volume={77},
  pages={451--550},
  year={1990},
  publisher={Wiley Online Library}
}

@article{holovko1990electrostatic,
  title={Electrostatic and packing contributions to the structure of water and aqueous electrolyte solutions},
  author={Holovko, MF and Kalyuzhny, Yu V and Heinzinger, K},
  journal={Zeitschrift fur Naturforschung A},
  volume={45},
  number={5},
  pages={687--694},
  year={1990},
  publisher={Verlag der Zeitschrift fur Naturforschung}
}

@article{smith1970analytical,
  title={Analytical representation of the Percus-Yevick hard-sphere radial distribution function},
  author={Smith, WR and Henderson, D},
  journal={Molecular Physics},
  volume={19},
  number={3},
  pages={411--415},
  year={1970},
  publisher={Taylor \& Francis}
}

@article{wertheim1986fluids,
  title={Fluids of dimerizing hard spheres, and fluid mixtures of hard spheres and dispheres},
  author={Wertheim, MS},
  journal={The Journal of chemical physics},
  volume={85},
  number={5},
  pages={2929--2936},
  year={1986},
  publisher={American Institute of Physics}
}

@book{hansen2013theory,
  title={Theory of simple liquids: with applications to soft matter},
  author={Hansen, Jean-Pierre and McDonald, Ian Ranald},
  year={2013},
  publisher={Academic press}
}

@book{gray2011theory,
  title={Theory of Molecular Fluids: Volume 1: Fundamentals},
  author={Gray, Christopher G and Gubbins, Keith E},
  volume={1},
  year={1984},
  publisher={Oxford University Press}
}

@article{zhang2003tethered,
  title={Tethered nano building blocks: Toward a conceptual framework for nanoparticle self-assembly},
  author={Zhang and Horsch, Mark A and Lamm, Monica H and Glotzer, Sharon C},
  journal={Nano letters},
  volume={3},
  number={10},
  pages={1341--1346},
  year={2003},
  publisher={ACS Publications}
}

@article{zhang2004self,
  title={Self-assembly of patchy particles},
  author={Zhang, Zhenli and Glotzer, Sharon C},
  journal={Nano letters},
  volume={4},
  number={8},
  pages={1407--1413},
  year={2004},
  publisher={ACS Publications}
}

@article{costa2005structure,
  title={Structure and equation of state of interaction site models for disc-shaped lamellar colloids},
  author={Costa*, Dino and Hansen, Jean-Pierre and Harnau, Ludger},
  journal={Molecular Physics},
  volume={103},
  number={14},
  pages={1917--1927},
  year={2005},
  publisher={Taylor \& Francis}
}

@article{delhorme2012monte,
  title={Monte Carlo simulations of a clay inspired model suspension: the role of rim charge},
  author={Delhorme, Maxime and J{\"o}nsson, Bo and Labbez, Christophe},
  journal={Soft Matter},
  volume={8},
  number={37},
  pages={9691--9704},
  year={2012},
  publisher={Royal Society of Chemistry}
}

@article{ruzicka2011observation,
  title={Observation of empty liquids and equilibrium gels in a colloidal clay},
  author={Ruzicka, Barbara and Zaccarelli, Emanuela and Zulian, Laura and Angelini, Roberta and Sztucki, Michael and Moussa{\"\i}d, Abdellatif and Narayanan, Theyencheri and Sciortino, Francesco},
  journal={Nature materials},
  volume={10},
  number={1},
  pages={56--60},
  year={2011},
  publisher={Nature Publishing Group UK London}
}

@article{raineri2001dielectrically,
  title={Dielectrically nontrivial closures for the RISM integral equation},
  author={Raineri, Fernando O and Stell, George},
  journal={The Journal of Physical Chemistry B},
  volume={105},
  number={47},
  pages={11880--11892},
  year={2001},
  publisher={ACS Publications}
}

@article{tavares2010percolation,
  title={Percolation of colloids with distinct interaction sites},
  author={Tavares, JM and Teixeira, PIC and Telo da Gama, MM},
  journal={Physical Review E—Statistical, Nonlinear, and Soft Matter Physics},
  volume={81},
  number={1},
  pages={010501},
  year={2010},
  publisher={APS}
}

@article{bianchi2008theoretical,
  title={Theoretical and numerical study of the phase diagram of patchy colloids: Ordered and disordered patch arrangements},
  author={Bianchi, Emanuela and Tartaglia, Piero and Zaccarelli, Emanuela and Sciortino, Francesco},
  journal={The Journal of chemical physics},
  volume={128},
  number={14},
  year={2008},
  publisher={AIP Publishing}
}

@article{ruzicka2011fresh,
  title={A fresh look at the Laponite phase diagram},
  author={Ruzicka, Barbara and Zaccarelli, Emanuela},
  journal={Soft Matter},
  volume={7},
  number={4},
  pages={1268--1286},
  year={2011},
  publisher={Royal Society of Chemistry}
}

\end{document}